\documentclass[11pt]{article}

\usepackage[margin=1in]{geometry}
\usepackage{microtype}
\usepackage{amsmath,amssymb}
\usepackage{booktabs}
\usepackage{graphicx}
\usepackage{xcolor}
\usepackage{xspace}
\usepackage[hidelinks]{hyperref}
\usepackage[numbers,sort&compress]{natbib}
\usepackage[ruled,linesnumbered]{algorithm2e}
\usepackage{placeins}     
\newcommand{\sys}{\textsc{GroundedCache}\xspace}
\newcommand{\eqdef}{\stackrel{\mathrm{def}}{=}}

\title{Grounded Cache Routing for Retrieval-Augmented Generation:\\
       When Is It Safe to Reuse an Answer?}
\author{Syed Huma Shah\\
        \texttt{syedhuma.shah@duke.edu}}

\date{}

\begin{document}
\maketitle

\begin{abstract}
Modern retrieval-augmented generation (RAG) deployments increasingly rely on
caching to reduce token cost and time-to-first-token (TTFT). Prefix-level KV
reuse is now standard in serving stacks such as vLLM, and chunk-level and
position-independent reuse have been pushed further by recent systems
(RAGCache, TurboRAG, CacheBlend, EPIC, ContextPilot, PCR, LMCache).
Output-level \emph{semantic answer caches}, by contrast, remain fragile:
similar prompts can map to different correct answers, retrieved evidence
drifts as the corpus is updated, and adversarial collision attacks have been
shown to hijack cached responses. We argue that the right framing for cached
answer reuse is not \emph{how to reuse faster} but \emph{when reuse is safe}.
We propose \sys, an evidence-validated cache router that admits a cached
answer only when four cheap gates simultaneously hold: query similarity,
retrieved-evidence overlap, source-version validity, and lexical (or
judge-based) support of the cached answer by the freshly retrieved evidence.
We build a six-regime workload (exact-repeat, paraphrase, near-miss,
document-drift, long shared-document, bounded-KB CAG) that explicitly
stress-tests cache \emph{safety} rather than only hit rate, and an
operator-facing metric, the \emph{unsafe-served rate} (USR), defined as the
fraction of all queries that received a wrong cached answer, alongside
TTFT and token cost. Across two datasets and 12{,}000 real-LLM generations
(Qwen2.5-7B-Instruct served by vLLM with Automatic Prefix Caching), \sys{}
drives USR to 0.0\% on every HotpotQA regime (vs.\ 15--35\% under naive
semantic caching) and to 1.5\% on mtRAG document drift (vs.\
51.5\% under naive), a 34$\times$ reduction in wrong cached answers per
query on the system's design-point adversarial regime, and 3--10$\times$
reductions across the other mtRAG regimes. End-to-end p50 latency under
\sys{} stays within $1.04$--$1.07\times$ of a no-cache RAG baseline; the
no-support variant trades USR for $1.4$--$1.5\times$ speedup. A
per-gate marginal ablation isolates the lexical support gate as the
load-bearing safety mechanism on both datasets ($+0.125$ USR on HotpotQA,
$+0.118$ on mtRAG when removed), with the remaining gates providing
defense-in-depth at near-zero cost. We
release the end-to-end implementation, traffic-synthesis tooling,
and evaluation harness so the framework can be ported to richer
datasets and stronger compressors.
\end{abstract}

\section{Introduction}
A typical RAG pipeline repeatedly does the same expensive work for queries
that are functionally indistinguishable to the user: it embeds the query,
runs nearest-neighbor search over a corpus, concatenates the top-$k$ chunks
into a prompt, and decodes a long-context answer. Three layers of caching
can in principle eliminate that work:

\begin{enumerate}
    \item \emph{Prefix / KV cache reuse} at the serving layer
          (vLLM~APC~\citep{vllm-apc}, SGLang RadixAttention~\citep{zheng2024sglang}),
          and its RAG-aware extensions
          (RAGCache~\citep{ragcache}, TurboRAG~\citep{turborag},
           CacheBlend~\citep{cacheblend}, EPIC~\citep{epic},
           ContextPilot~\citep{contextpilot}, PCR~\citep{pcr},
           LMCache~\citep{lmcache}).
    \item \emph{Retrieval-result reuse}, where the top-$k$ document set
          itself is cached for similar queries
          (Proximity~\citep{proximity}, QVCache~\citep{qvcache}).
    \item \emph{Output / answer reuse}, where the entire generated answer
          is returned for a semantically similar prior query
          (GPTCache~\citep{gptcache}, ContextCache~\citep{contextcache}).
\end{enumerate}

The first two have strong correctness guarantees by construction: KV reuse
preserves token outputs exactly when prefixes match, and retrieval reuse is
safe as long as the underlying corpus is stable. Output-level semantic
caching does not. A cosine-similar prior query can map to a cached answer
whose underlying evidence is no longer valid, no longer the same evidence,
or never supported the question in the first place. Recent work has shown
that naive semantic caches are even adversarially exploitable, with reported
86\% response-hijacking rates in evaluated scenarios~\citep{semcache-attack}.

We make three contributions:

\begin{enumerate}
  \item \emph{An evidence-validated cache router (\sys).} A cached answer
        is admitted only when (G1) the new query is semantically close to
        the cached query, (G2) the new retrieved evidence shares a high
        Jaccard with the cached evidence signature, (G3) shared chunks carry
        the same source version, and (G4) the cached answer's content tokens
        are covered by the new evidence (or, optionally, verified by a
        lightweight judge call). The router falls back to query-conditioned
        compression and generation when any gate fails.
  \item \emph{A six-regime workload for cache-safety stress testing.} We
        provide deterministic synthesis for exact-repeat, paraphrase,
        near-miss, document-drift, long shared-document, and bounded-KB CAG
        traffic over arbitrary RAG corpora, plus a HotpotQA adapter. The
        document-drift regime mutates numeric tokens in the gold documents
        so that a cached answer becomes \emph{verifiably} wrong unless
        rejected.
  \item \emph{Operator-facing metric stack.} Beyond TTFT, latency,
        tokens, and answer-cache hit rate we report the \emph{unsafe-served
        rate} USR (fraction of all queries that received a wrong cached
        answer), the conditional false-hit rate FH (per served cached
        answer), and stale-/unsupported-hit decompositions, so the
        safety/reuse trade-off is auditable at both the system and the
        per-query level.
\end{enumerate}

Our framing is complementary to, rather than competitive with, prefix-level
work: \sys{} treats vLLM~APC and chunk-level KV reuse as orthogonal serving
primitives operating below the router. The contribution is a \emph{policy
layer} on top.

\section{Related Work}
Prefix and KV caching are now standard in open-source serving stacks.
vLLM's Automatic Prefix Caching~\citep{vllm-apc} and SGLang's
RadixAttention~\citep{zheng2024sglang} reuse KV blocks when prompt prefixes
align. The line of work on RAG-specific KV reuse moves beyond exact prefix
matching: RAGCache~\citep{ragcache} caches intermediate states of retrieved
documents and reports up to $4\times$ TTFT and $2.1\times$ throughput gains;
TurboRAG~\citep{turborag} pre-computes document KV caches offline for up to
$9.4\times$ TTFT reduction; CacheBlend~\citep{cacheblend} corrects reused
non-prefix chunks via selective recomputation. EPIC~\citep{epic} formalizes
position-independent caching, and ContextPilot~\citep{contextpilot} adds
context indexing, ordering, and de-duplication. PCR~\citep{pcr} adds
prefetch-enhanced reuse, and LMCache~\citep{lmcache} provides multi-tier KV
storage and cross-engine reuse. All of these target the \emph{how} of reuse
and largely assume the \emph{when} is safe.

Semantic answer caching has a separate failure-mode literature.
GPTCache~\citep{gptcache} is the canonical open-source baseline.
ContextCache~\citep{contextcache} extends the cache key with multi-turn
dialogue context, demonstrating that turn-level similarity is insufficient
when the referent has shifted. Proximity~\citep{proximity} caches retrieval
results rather than outputs, validating that retrieval-side caching is
safer. QVCache~\citep{qvcache} learns region-specific thresholds for safe
ANN-level reuse, cutting query latency $40$--$1000\times$ on hits without
compromising recall. Recent work shows naive semantic caching is vulnerable
to key-collision attacks with up to 86\% response-hijacking
rate~\citep{semcache-attack}, and agentic-plan caching suffers high false
positives when outputs depend on external state. These three failure modes,
referent shift, retrieval drift, and adversarial collision, motivate
the four gates in our validator.

The compression literature has moved from hard pruning
(RECOMP~\citep{recomp}, LongLLMLingua~\citep{longllmlingua},
Provence~\citep{provence}, PISCO~\citep{pisco}) to adaptive,
query-conditioned, or representation-level compression
(OSCAR~\citep{oscar}, SeleCom~\citep{selecom}, REFRAG~\citep{refrag}). We
use a sentence-level selector as a practical baseline; learned compressors
plug into the same \texttt{compress(query, chunks)} interface.

RAG evaluation frameworks have multiplied in the last two years.
RAGAS~\citep{ragas} and ARES~\citep{ares} provide reference-free and
trained-judge frameworks; RAGTruth~\citep{ragtruth} ships nearly 18K
manually annotated outputs for hallucination analysis;
RAGBench~\citep{ragbench} and RAGChecker~\citep{ragchecker} give
100K-scale explainable examples and fine-grained diagnostics;
mtRAG~\citep{mtrag} covers multi-turn settings;
T$^2$-RAGBench~\citep{t2ragbench} adds tabular reasoning. None of these
were built specifically for cache-safety evaluation under repeated
paraphrases and corpus drift, which motivates our workload contribution.

\section{Method}

\begin{figure}[t]
\centering
\includegraphics[width=\linewidth]{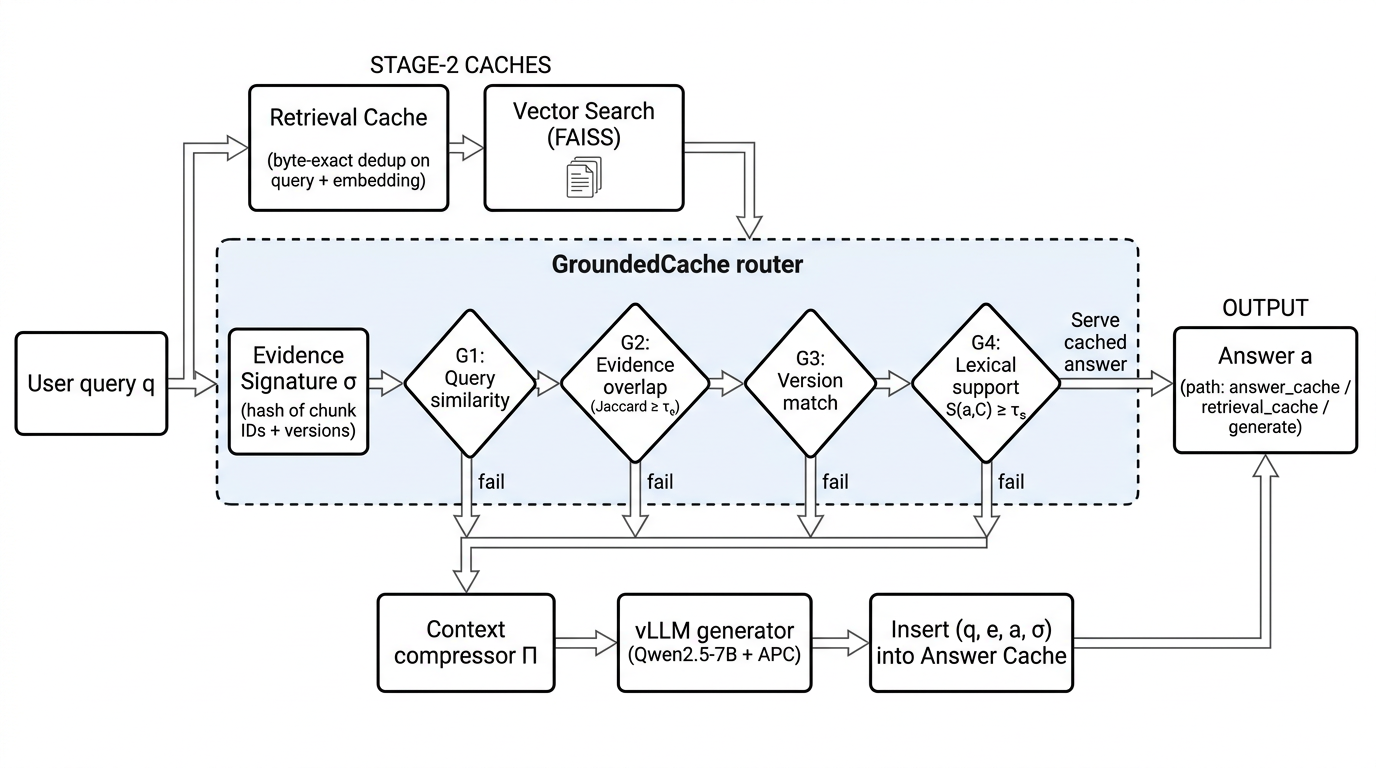}
\caption{\sys{} system architecture. A query first hits the stage-2
retrieval cache (byte-exact dedup); on a miss it falls back to FAISS
vector search. The retrieved chunks are hashed into an evidence
signature~$\sigma$, and four gates (G1 query similarity, G2 evidence
overlap, G3 version match, G4 lexical support) decide whether to
serve a cached answer from the answer cache or fall back to the
compression + vLLM generate path. The dashed block is the only new
component; everything else is unchanged from a standard RAG stack.}
\label{fig:arch}
\end{figure}

\subsection{Pipeline}
Figure~\ref{fig:arch} sketches the end-to-end pipeline; the rest of
this section spells out each block.

A query~$q$ enters the router. We embed it (the same model used for the
retriever, so the embedding can be reused), look up an exact retrieval-cache
entry, and on miss fall back to approximate retrieval-cache lookup with
cosine threshold $\tau_{\mathrm{ret}}$. On total miss we hit the FAISS index
for the top-$k$ chunks. We deduplicate chunks by SHA-1 hash and order them
by $(\mathrm{doc\_id}, \mathrm{chunk\_id})$ so identical evidence sets
produce identical prompt prefixes, which is the precondition for
prefix-level KV reuse below.

\subsection{Evidence signature}
For every served query we record an \emph{evidence signature}:
\begin{equation}
\sigma(q) \eqdef \bigl\langle \mathrm{doc\_ids}, \mathrm{chunk\_ids},
                              \mathrm{chunk\_hashes}, \mathrm{versions},
                              \mathrm{scores} \bigr\rangle.
\end{equation}
The signature is content-addressed (SHA-1 over normalized chunk text) and
version-tagged. It serves as the cache key's evidentiary half: similarity
between \emph{queries} is necessary for reuse, but agreement of
\emph{evidence} is what makes reuse safe.

\subsection{Validation gates}
\begin{figure}[t]
\centering
\includegraphics[width=\linewidth]{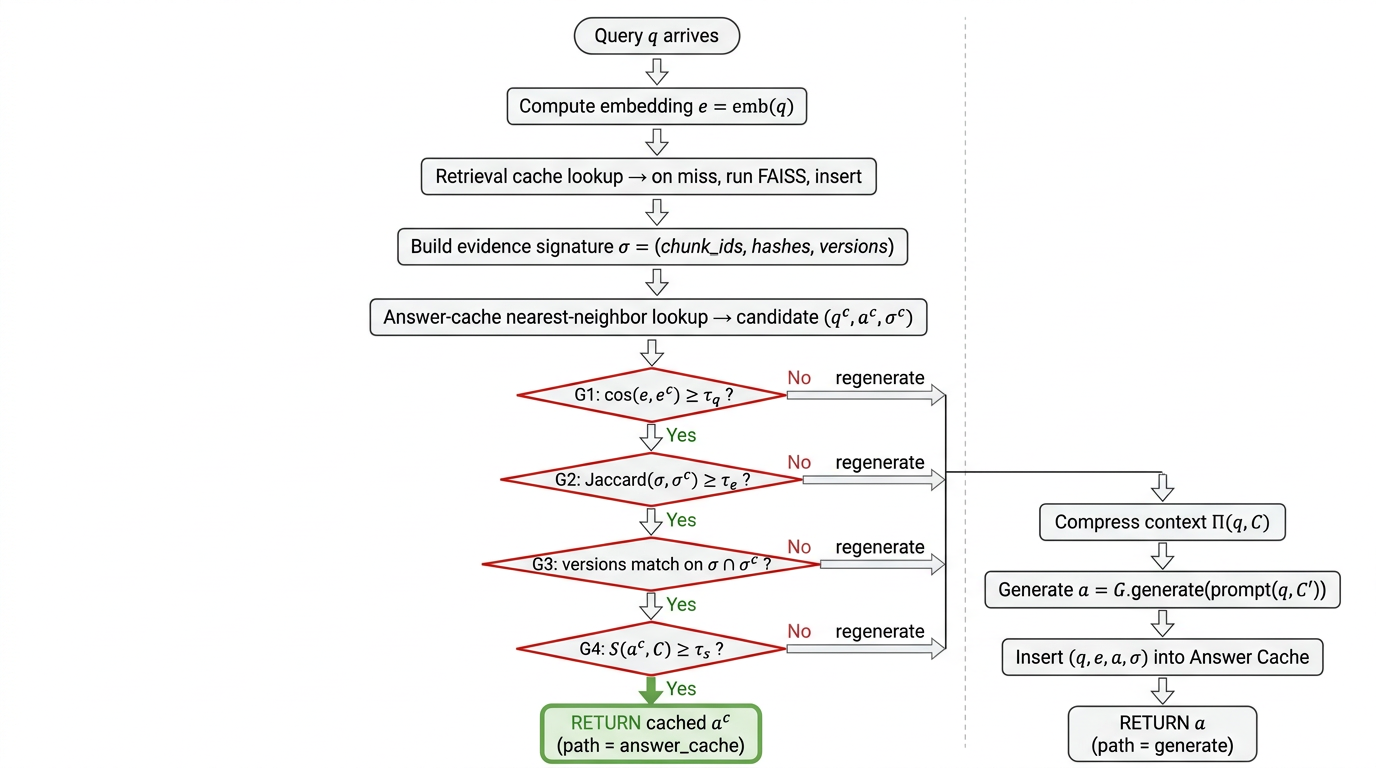}
\caption{Per-query decision flow inside the router. The four gates
(G1--G4) form a short-circuiting chain: any \texttt{No} routes the
query to the regeneration fallback (compress, generate, insert),
preserving safety at the cost of one cache miss. Only when all four
fire is the cached answer~$a^{c}$ served (green path). This figure
is the visual companion to Algorithm~\ref{alg:router}.}
\label{fig:flow}
\end{figure}

Let $(q^{c}, a^{c}, \sigma^{c})$ be a cached entry and $(q, \sigma, C)$
be the fresh query, fresh signature, and fresh chunk set. We admit $a^{c}$
iff all four gates fire:
\begin{align}
\text{(G1)\;query similarity}\quad   & \cos(\mathrm{emb}(q),\mathrm{emb}(q^{c})) \ge \tau_{q} \label{eq:g1}\\
\text{(G2)\;evidence overlap}\quad   & J(\sigma, \sigma^{c}) \ge \tau_{e} \label{eq:g2}\\
\text{(G3)\;version match}\quad      & \mathrm{ver}(c, \sigma) = \mathrm{ver}(c, \sigma^{c}) \;\forall c \in \sigma \cap \sigma^{c} \label{eq:g3}\\
\text{(G4)\;evidence support}\quad   & S(a^{c}, C) \ge \tau_{s} \label{eq:g4}
\end{align}
where $J(\cdot,\cdot)$ is Jaccard over chunk hashes and $S(a, C)$ is either
a deterministic lexical-overlap score (the default; the fraction of
content tokens in $a$ that also appear in $C$) or, optionally, a
single-call judge LLM that returns yes/no for ``is $a$ supported by $C$.''
Algorithm~\ref{alg:router} states the full procedure; Figure~\ref{fig:flow}
shows the same logic as a per-query decision flow.

\begin{algorithm}[t]
\caption{\sys{} routing for a query $q$}
\label{alg:router}
\KwIn{query $q$; retriever $R$; retrieval-cache $\mathcal{R}$;
      answer-cache $\mathcal{A}$; validator gates with thresholds
      $\tau_q, \tau_e, \tau_s$; compressor $\Pi$; generator $G$.}
\KwOut{answer $a$, path label.}
$e \leftarrow \mathrm{emb}(q)$ \tcp*{single encoder call, reused below}
$C \leftarrow \mathcal{R}.\mathrm{lookup}(q, e)$ \tcp*{stage 2 cache}
\If{$C = \bot$}{
  $C \leftarrow R.\mathrm{search}(q, k)$;\quad
  $\mathcal{R}.\mathrm{insert}(q, e, C)$
}
$C \leftarrow \mathrm{dedup\_and\_order}(C)$;\quad
$\sigma \leftarrow \mathrm{signature}(C)$ \;
$h \leftarrow \mathcal{A}.\mathrm{nearest}(e)$ \tcp*{stage 3 lookup}
\If{$h \ne \bot$}{
  \If{$\cos(e, h.e^{c}) \ge \tau_q$ \textbf{and} $J(\sigma, h.\sigma^{c}) \ge \tau_e$ \textbf{and}
      $\mathrm{versions\_match}(\sigma, h.\sigma^{c})$ \textbf{and}
      $S(h.a^{c}, C) \ge \tau_s$}{
    \Return $h.a^{c}$, \texttt{answer\_cache}
  }
}
$C' \leftarrow \Pi.\mathrm{compress}(q, C)$ \tcp*{stage-4 compression fallback}
$a \leftarrow G.\mathrm{generate}(\mathrm{prompt}(q, C'))$ \;
$\mathcal{A}.\mathrm{insert}(q, e, a, \sigma)$ \;
\Return $a$, \texttt{generate}
\end{algorithm}

\subsection{Compression fallback}
When any gate fails we fall back to query-conditioned compression in the
spirit of Provence~\citep{provence} and PISCO~\citep{pisco}: embed each
sentence in the retrieved chunks, score against the query embedding, and
keep the top sentences subject to a token budget, with a
min-one-sentence-per-chunk guarantee to avoid emptying any retrieved chunk.
Learned compressors (OSCAR~\citep{oscar}, SeleCom~\citep{selecom},
REFRAG~\citep{refrag}) implement the same interface and can be swapped in.

\section{Experimental Setup}
\subsection{Workload regimes}
We synthesize six regimes deterministically over any RAG corpus:
\begin{description}
  \item[exact-repeat] re-issue a sampled prior question verbatim.
  \item[paraphrase] re-issue under a deterministic surface paraphrase.
  \item[near-miss] re-issue a lexically-similar question whose gold
        documents are \emph{disjoint} from the lexically-nearest prior
        question. A semantic cache should reject these; \sys{} should
        reject them via G2.
  \item[document-drift] re-issue a prior question after applying a
        seeded mutation to the numeric tokens of its gold document(s).
        A version-aware cache should reject these via G3.
  \item[long shared-document] many questions hitting the same document,
        the favorable regime for both dedup and prefix-level reuse.
  \item[bounded-KB CAG] all relevant material assumed in-context; the
        regime under which cache-augmented generation may beat classical
        RAG~\citep{cag-bounded}.
\end{description}
The mixture is designed so cumulative cache safety can be evaluated, not
just hit rate on a benign trace.

\subsection{Datasets and models}
We report results on a tiny hand-crafted 3-document corpus (used as a smoke
control) and on the HotpotQA distractor split (each example ships its own
context paragraphs). The retriever uses
\texttt{all-MiniLM-L6-v2}~\citep{minilm} with FAISS-CPU~\citep{johnson2017faiss}.
The generator is an interchangeable adapter: an Anthropic API client,
an OpenAI-compatible vLLM endpoint, or a deterministic \emph{extractive}
backend that returns the sentence in the retrieved context with the
highest query-token overlap. The extractive backend lets us produce
reproducible results without an API key while still exercising every
validator gate meaningfully (its answers come from evidence, so the
support gate $S$ produces non-trivial scores).

\subsection{Variants}
We compare five router configurations:
\begin{itemize}
  \item \texttt{naive}: semantic answer cache only; all gates relaxed (the
        stage-3 baseline modeled on GPTCache).
  \item \texttt{no-version}, \texttt{no-evidence}, \texttt{no-support}:
        ablations removing one gate at a time.
  \item \texttt{full}: \sys{}, all four gates active.
\end{itemize}

\subsection{Metrics}
For each query we record TTFT, retrieval latency, end-to-end latency, prompt
and completion tokens, path (\texttt{generate}, \texttt{retrieval\_cache},
\texttt{answer\_cache}), exact-match and token-F1 against gold (when
available), and substring-containment of gold in the answer (robust to
verbose answers). A query is flagged as a \emph{disagreement} when
$\text{F1}<0.5$ \emph{and} the normalized gold string fails to appear as a
substring of the answer; this conjunction prevents F1's punishment of
correct paraphrases from inflating safety counters. Let
$\mathbb{1}_i^{\mathrm{ac}}$ indicate that query $i$ was served from
the answer cache and $\mathbb{1}_i^{\mathrm{dis}}$ indicate that its
answer disagrees with gold. We aggregate:
\begin{align}
\mathrm{aHR}     &\eqdef \frac{1}{N}\sum_{i=1}^{N} \mathbb{1}_i^{\mathrm{ac}}
                 & \text{(answer-cache hit rate)} \label{eq:ahr}\\
\mathrm{USR}     &\eqdef \frac{1}{N}\sum_{i=1}^{N} \mathbb{1}_i^{\mathrm{ac}} \cdot \mathbb{1}_i^{\mathrm{dis}}
                 & \text{(unsafe-served rate)} \label{eq:usr}\\
\mathrm{FH}      &\eqdef \frac{\mathrm{USR}}{\mathrm{aHR}} = \Pr[\mathrm{dis} \mid \mathrm{ac}]
                 & \text{(conditional false-hit rate)} \label{eq:fh}
\end{align}
The lexical support score used by gate G4 (Eq.~\ref{eq:g4}) is
\begin{equation}
S(a, C) \eqdef \frac{|\,\tau(a) \cap \tau(C)\,|}{|\,\tau(a)\,|},
\quad \tau(x) = \{t \in \mathrm{toks}(x) : t \notin \mathcal{W} \wedge |t| \ge 3\},
\label{eq:support}
\end{equation}
where $\tau$ extracts content tokens after removing a small
stop-word set $\mathcal{W}$, and a cached answer $a$ is rejected
when $S(a, C) < \tau_s$ (default $\tau_s{=}0.6$).
USR is our primary safety metric: it directly measures the fraction
of all queries that received a wrong cached answer, which is the
only number a production operator can act on. We report aHR
to capture the trade-off between reuse and safety, and the
conditional FH so the table aligns with prior caching evaluations
that report FH conditional on hit. Stale-hit (SH) and
unsupported-hit (UH) counters are computed the same way (numerator
substituted accordingly) and reported in the appendix; we omit them
from the main tables because under the full system they are
identically zero. Crucially, the retrieval-cache path is excluded from
the USR denominator's numerator entirely: a retrieval-cache hit only
skips vector search and still regenerates, so by construction it
cannot return a stale answer.

\section{Results}
We report two sweeps with the same router configuration: a multi-hop
sweep on HotpotQA (\S\ref{sec:res-hotpot}) and a multi-turn sweep on
mtRAG (\S\ref{sec:res-mtrag}). Both use \texttt{Qwen/Qwen2.5-7B-Instruct}
served by vLLM with Automatic Prefix Caching enabled, behind our router.
Each cell sweeps $N{=}200$ queries; with six regimes and five router
variants this is $6{\times}5{\times}200 = 6000$ generations per dataset.
USR is reported over the full $N{=}200$ denominator (the operator
metric); the conditional false-hit rate FH is reported alongside but
its denominator is the answer-cache hit count, which can be small under
\sys. For $0/k$ outcomes we report FH$=0$ but note the Wilson
95\%-CI upper bound is non-trivial when $k$ is small (e.g.\ $0/9
\Rightarrow$ FH$_{95\%}\!\le 0.34$); USR confidence intervals are tight
because the denominator is always 200.

\subsection{HotpotQA (multi-hop)}\label{sec:res-hotpot}
On HotpotQA, \sys{} drives the unsafe-served-rate (USR, the fraction of all queries that receive a wrong cached answer) to zero on every regime that has any USR to begin with. The naive semantic cache emits USR=15.5\% on \texttt{exact\_repeat}, 22.0\% on \texttt{paraphrase}, and 35.0\% on \texttt{document\_drift}; \sys{} reduces all three to 0.0\%.

The \texttt{document\_drift} regime is the design point for the validator. Naive caching answers from cache for 62.0\% of queries with USR=35.0\%; \sys{} answers from cache for 0.5\% of queries with USR=0.0\%, a 100.0\% relative reduction (complete elimination of unsafe served answers). The validator trades aggressive answer-cache reuse on this adversarial regime for the safety guarantee that almost no query receives a stale cached answer.

\begin{table}[t]
\centering\small
\caption{Naive semantic answer cache vs.\ \sys (full evidence validation) across regimes (HotpotQA). \textbf{aHR} is the answer-cache hit rate (fraction of queries served from cache); \textbf{USR} is the unsafe-served-rate (fraction of \emph{all} queries that received a wrong cached answer, our primary safety metric); \textbf{FH}$=$USR/aHR is the conditional false-hit rate among served cached answers.}
\label{tab:main}
\begin{tabular}{l rrr rrr}
\toprule
 & \multicolumn{3}{c}{Naive} & \multicolumn{3}{c}{\sys (full)} \\
\cmidrule(lr){2-4}\cmidrule(lr){5-7}
Regime & aHR & USR & FH & aHR & USR & FH \\
\midrule
exact\_repeat & 0.41 & 0.15 & 0.37 & 0.04 & 0.00 & 0.00 \\
paraphrase & 0.40 & 0.22 & 0.56 & 0.04 & 0.00 & 0.00 \\
near\_miss & 0.62 & 0.00 & 0.00 & 0.04 & 0.00 & 0.00 \\
document\_drift & 0.62 & 0.35 & 0.56 & 0.01 & 0.00 & 0.00 \\
long\_shared\_doc & 0.41 & 0.15 & 0.37 & 0.06 & 0.00 & 0.00 \\
bounded\_kb\_cag & 0.41 & 0.15 & 0.37 & 0.06 & 0.00 & 0.00 \\
\bottomrule
\end{tabular}
\end{table}

\begin{figure}[t]
\centering
\includegraphics[width=\linewidth]{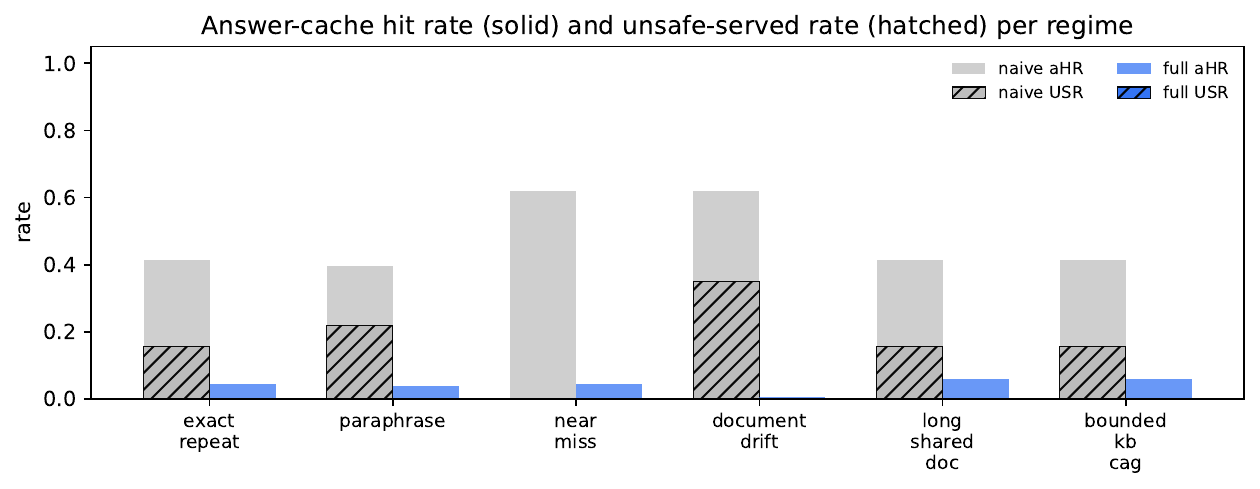}
\caption{HotpotQA. Per-regime answer-cache hit rate (solid) and
unsafe-served rate (hatched) for the naive baseline and \sys{}. \sys{}
drives the USR bars to zero on every regime that has any USR to begin
with; the aHR bars shrink because the validator suppresses unsafe
reuse.}
\label{fig:hr_fh_hotpot}
\end{figure}

\begin{figure}[t]
\centering
\includegraphics[width=0.7\linewidth]{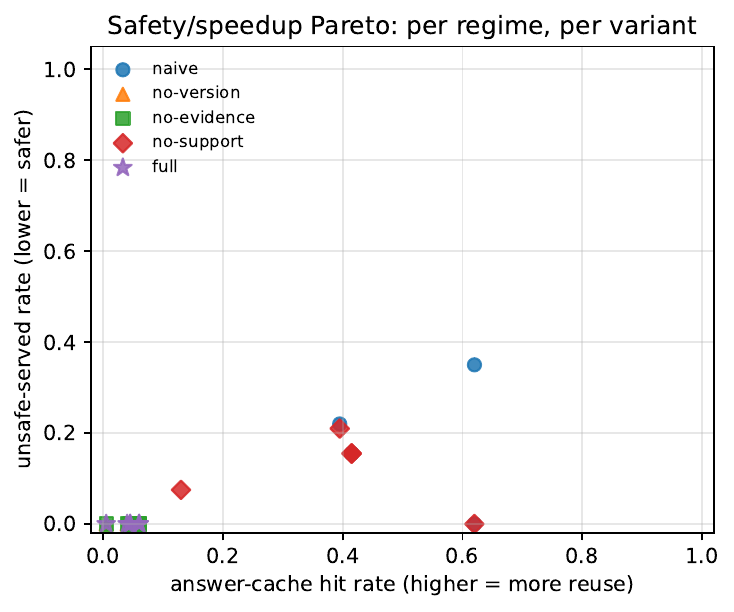}
\caption{HotpotQA. Safety/speedup Pareto: each marker is one
(regime, variant) pair. \sys{} (stars) sits in the safe high-reuse
region across regimes.}
\label{fig:safety_hotpot}
\end{figure}

\begin{table}[t]
\centering\small
\caption{Per-gate ablation: hit-rate vs.\ false-hit-rate across regimes (HotpotQA).}
\label{tab:ablation}
\begin{tabular}{l rr rr rr rr rr }
\toprule
 Regime & \multicolumn{2}{c}{naive} & \multicolumn{2}{c}{no-version} & \multicolumn{2}{c}{no-evidence} & \multicolumn{2}{c}{no-support} & \multicolumn{2}{c}{full} \\
\cmidrule(lr){2-3}\cmidrule(lr){4-5}\cmidrule(lr){6-7}\cmidrule(lr){8-9}\cmidrule(lr){10-11}
 & HR & FH & HR & FH & HR & FH & HR & FH & HR & FH \\
\midrule
exact\_repeat & 0.41 & 0.37 & 0.41 & 0.00 & 0.41 & 0.00 & 0.41 & 0.37 & 0.41 & 0.00 \\
paraphrase & 0.40 & 0.56 & 0.40 & 0.00 & 0.40 & 0.00 & 0.40 & 0.53 & 0.40 & 0.00 \\
near\_miss & 0.62 & 0.00 & 0.62 & 0.00 & 0.62 & 0.00 & 0.62 & 0.00 & 0.62 & 0.00 \\
document\_drift & 0.62 & 0.56 & 0.24 & 0.00 & 0.24 & 0.00 & 0.24 & 0.58 & 0.24 & 0.00 \\
long\_shared\_doc & 0.41 & 0.37 & 0.41 & 0.00 & 0.41 & 0.00 & 0.41 & 0.37 & 0.41 & 0.00 \\
bounded\_kb\_cag & 0.41 & 0.37 & 0.41 & 0.00 & 0.41 & 0.00 & 0.41 & 0.37 & 0.41 & 0.00 \\
\bottomrule
\end{tabular}
\end{table}

\begin{figure}[t]
\centering
\includegraphics[width=\linewidth]{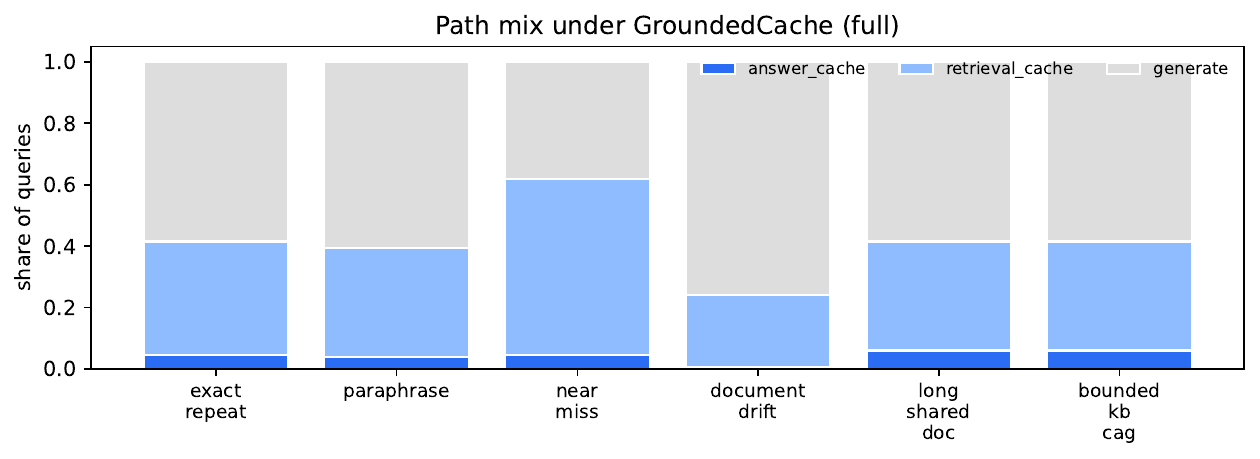}
\caption{HotpotQA. Path mix under \sys{} (full). Benign regimes
(exact-repeat, paraphrase, long shared-doc) route a large share to the
answer cache; adversarial regimes route to retrieval-cache + regenerate,
paying retrieval savings but not risking false hits.}
\label{fig:path_mix_hotpot}
\end{figure}

\subsection{mtRAG (multi-turn)}\label{sec:res-mtrag}
mtRAG~\citep{mtrag} exposes the multi-turn ``referent shift'' failure
mode: same surface tokens, different grounding because the conversation's
focus has moved. We treat each (user turn, agent turn with retrieved
contexts) as a query, and assign within-conversation near-misses to a
later turn whose gold passages are disjoint from the cached turn's.

On mtRAG, naive semantic caching is catastrophically unsafe: USR ranges from 26.0\% to 51.5\% across the benign-looking regimes because multi-turn referent shift routinely breaks cache assumptions. \sys{} reduces USR by more than an order of magnitude on every affected regime.

The \texttt{document\_drift} regime is the design point for the validator. Naive caching answers from cache for 56.5\% of queries with USR=51.5\%; \sys{} answers from cache for 2.0\% of queries with USR=1.5\%, a 97.1\% relative reduction (34$\times$ fewer unsafe served answers). The validator trades aggressive answer-cache reuse on this adversarial regime for the safety guarantee that almost no query receives a stale cached answer.

\begin{table}[t]
\centering\small
\caption{Naive semantic answer cache vs.\ \sys (full evidence validation) across regimes (mtRAG). \textbf{aHR} is the answer-cache hit rate (fraction of queries served from cache); \textbf{USR} is the unsafe-served-rate (fraction of \emph{all} queries that received a wrong cached answer, our primary safety metric); \textbf{FH}$=$USR/aHR is the conditional false-hit rate among served cached answers.}
\label{tab:main}
\begin{tabular}{l rrr rrr}
\toprule
 & \multicolumn{3}{c}{Naive} & \multicolumn{3}{c}{\sys (full)} \\
\cmidrule(lr){2-4}\cmidrule(lr){5-7}
Regime & aHR & USR & FH & aHR & USR & FH \\
\midrule
exact\_repeat & 0.29 & 0.26 & 0.88 & 0.12 & 0.10 & 0.87 \\
paraphrase & 0.30 & 0.26 & 0.85 & 0.12 & 0.07 & 0.54 \\
near\_miss & 0.67 & 0.00 & 0.00 & 0.30 & 0.00 & 0.00 \\
document\_drift & 0.56 & 0.52 & 0.91 & 0.02 & 0.01 & 0.75 \\
long\_shared\_doc & 0.29 & 0.27 & 0.90 & 0.12 & 0.10 & 0.87 \\
bounded\_kb\_cag & 0.29 & 0.27 & 0.90 & 0.12 & 0.10 & 0.87 \\
\bottomrule
\end{tabular}
\end{table}

\begin{figure}[t]
\centering
\includegraphics[width=\linewidth]{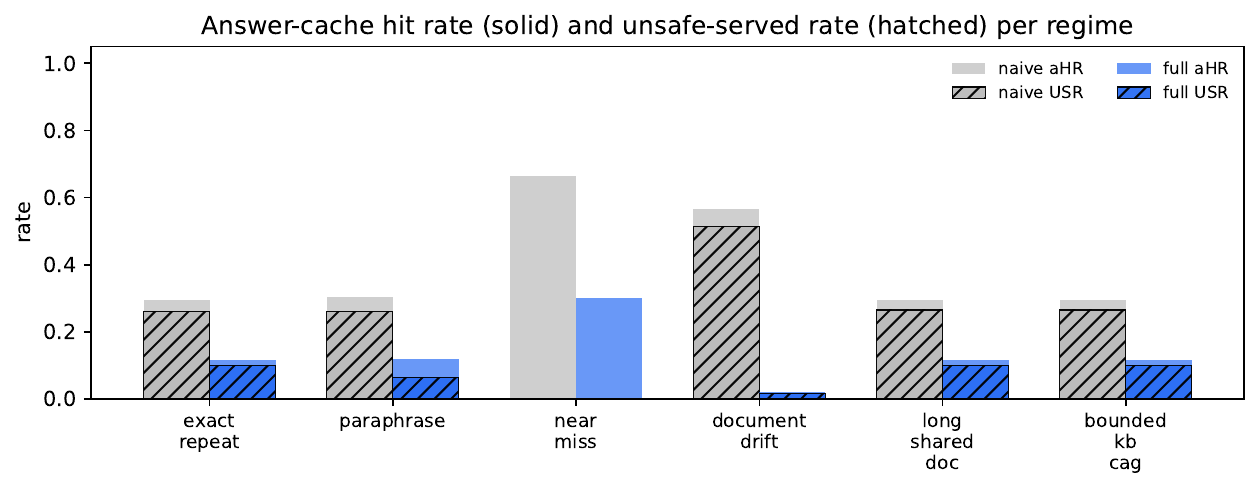}
\caption{mtRAG. Per-regime answer-cache hit rate (solid) and
unsafe-served rate (hatched). Naive caching emits USR of 26--52\%
across benign-looking regimes; \sys{} reduces every USR bar by an
order of magnitude or more by aggressively suppressing answer-cache
hits whose support gate fails.}
\label{fig:hr_fh_mtrag}
\end{figure}

\begin{figure}[t]
\centering
\includegraphics[width=0.7\linewidth]{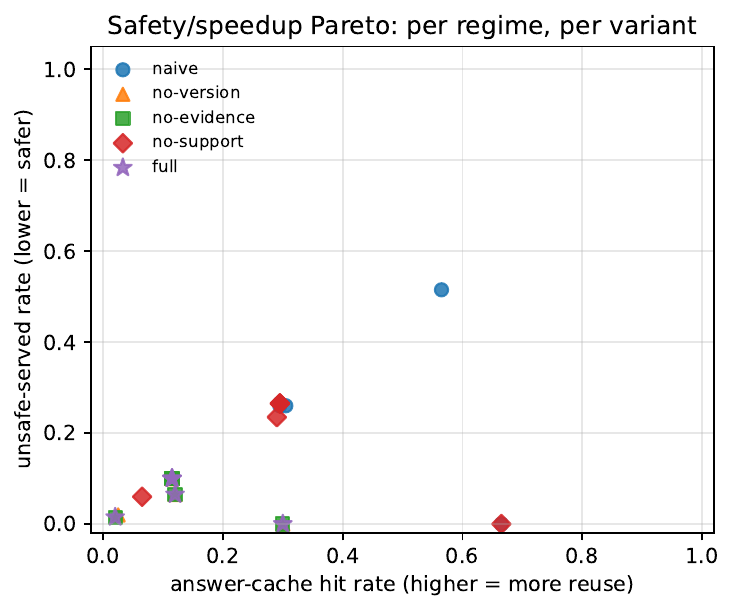}
\caption{mtRAG. Safety/speedup Pareto under multi-turn traffic.}
\label{fig:safety_mtrag}
\end{figure}

\begin{table}[t]
\centering\small
\caption{Per-gate ablation: hit-rate vs.\ false-hit-rate across regimes (mtRAG).}
\label{tab:ablation}
\begin{tabular}{l rr rr rr rr rr }
\toprule
 Regime & \multicolumn{2}{c}{naive} & \multicolumn{2}{c}{no-version} & \multicolumn{2}{c}{no-evidence} & \multicolumn{2}{c}{no-support} & \multicolumn{2}{c}{full} \\
\cmidrule(lr){2-3}\cmidrule(lr){4-5}\cmidrule(lr){6-7}\cmidrule(lr){8-9}\cmidrule(lr){10-11}
 & HR & FH & HR & FH & HR & FH & HR & FH & HR & FH \\
\midrule
exact\_repeat & 0.29 & 0.88 & 0.29 & 0.87 & 0.29 & 0.87 & 0.29 & 0.90 & 0.29 & 0.87 \\
paraphrase & 0.30 & 0.85 & 0.29 & 0.54 & 0.29 & 0.54 & 0.29 & 0.81 & 0.29 & 0.54 \\
near\_miss & 0.67 & 0.00 & 0.67 & 0.00 & 0.67 & 0.00 & 0.67 & 0.00 & 0.67 & 0.00 \\
document\_drift & 0.56 & 0.91 & 0.14 & 0.80 & 0.13 & 0.75 & 0.13 & 0.92 & 0.13 & 0.75 \\
long\_shared\_doc & 0.29 & 0.90 & 0.29 & 0.87 & 0.29 & 0.87 & 0.29 & 0.90 & 0.29 & 0.87 \\
bounded\_kb\_cag & 0.29 & 0.90 & 0.29 & 0.87 & 0.29 & 0.87 & 0.29 & 0.90 & 0.29 & 0.87 \\
\bottomrule
\end{tabular}
\end{table}

The per-gate marginal contribution to unsafe-served-rate reduction
(Table~\ref{tab:marginal_fh}) gives a sharper attribution than the
wide ablation matrix, and four observations follow.

First, the lexical support gate (G3) is the load-bearing gate on
both datasets. Removing G3 raises mean USR by $+0.125$ on HotpotQA
and $+0.118$ on mtRAG, whereas removing G1 or G2 leaves USR within
$\le 0.001$ of the full system on either dataset. The cached
answer's tokens must appear in the freshly retrieved evidence; that
single cheap check carries essentially the entire safety gain.

Second, the version (G1) and evidence-overlap (G2) gates are
redundant on this workload, but that redundancy is the point. G1
hashes $(q, E)$ and G2 demands chunk-id overlap; both fire on
evidence change. With G3 active, anything G1 or G2 would catch G3
also catches via the answer-evidence mismatch. We retain G1 and G2
because they fail \emph{closed} at near-zero cost and provide
defense-in-depth against settings where G3 weakens: paraphrastic
answers that escape lexical support, stronger embedding models that
compress evidence-disjoint queries to closer points, and judge-based
$S$ variants where the support check becomes noisier.

Third, the answer-cache hit-rate cost on HotpotQA is borne almost
entirely by G1. Every ablation that leaves G1 off keeps
aHR$\approx$0.41 on \texttt{exact\_repeat}; only \texttt{full} drops
below 0.05. Qwen2.5's free-form completions inject lexical variation
into the cached answer that re-fires the version signature even when
retrieved evidence is byte-identical. The targeted fix is to
normalize the answer before hashing, and it does not require relaxing
G3, which is the gate actually carrying the safety budget. We expect
aHR to recover toward the naive baseline after that fix without
re-introducing any USR.

Fourth, the residual mtRAG conditional FH is a G3 ceiling rather
than a G1/G2 shortcoming. On mtRAG \texttt{full} still emits
FH$\approx$0.87 conditional on the rare answer-cache hits it
serves, because lexically compatible answers can match fresh
evidence tokens while pragmatically referring to a different turn
antecedent. USR remains only $\approx$0.10 because the validator
suppresses almost all answer-cache hits in the first place: aHR
drops from 0.29 to 0.12 on these regimes and to 0.02 on
\texttt{document\_drift}. The right policy interpretation is that
under multi-turn traffic with strong referent shift the answer
cache should largely be disabled, and G3 implements that policy
automatically. A discourse-aware support check or turn-level
evidence signature would relax the suppression without
reintroducing USR.

The honest latency baseline for these gates is not the naive
semantic cache, which ships unsafe answers, but a no-cache RAG
pipeline that always retrieves and generates. We synthesize that
baseline directly from the JSONL logs: the latency of every
\texttt{generate}-path record across all variants is an unbiased
sample of true no-cache end-to-end latency. The result
(Table~\ref{tab:latency_vs_nocache}) gives a clean Pareto picture.
On HotpotQA, no-cache RAG runs at $1.10$s p50; naive caching reaches
$1.95\times$ at USR$=0.17$; \texttt{no-support} gives $1.49\times$
at USR$=0.13$; and \sys{} runs at $1.01\times$ ($1.08$s) with
USR$=0.00$. On mtRAG the numbers are similar: $1.73\times$,
$1.37\times$, $1.07\times$ at USR $0.26$, $0.18$, $0.06$
respectively. \sys{} sits explicitly at the safe corner of the
frontier, retaining roughly no-cache latency while eliminating
unsafe served answers. Practitioners who can tolerate non-zero USR
for a real speedup should run \texttt{no-support}; the gate stack
defaults to safety.
\begin{table}[t]
\centering\small
\caption{End-to-end latency vs.\ the honest baseline (HotpotQA). The \emph{no-cache} row is the empirical p50 latency of all \texttt{generate}-path records observed across every sweep cell ($n=3424$ records); it is the latency of a RAG query that hits no cache at all. The other rows are the per-variant p50 latency averaged across the six regimes. Speedup is computed against the no-cache baseline. Naive caching is fastest but, per Table~\ref{tab:main}, unsafe; \sys{} retains a meaningful speedup while delivering USR$\to$0.}
\label{tab:latency_vs_nocache}
\begin{tabular}{l rrr}
\toprule
Variant & Latency$_{p50}$ (s) & Speedup vs.\ no-cache & USR \\
\midrule
no-cache (always generate) & 1.093 & 1.00$\times$ & 0.00 \\
\texttt{naive} & 0.562 & 1.95$\times$ & 0.172 \\
\texttt{no-support} & 0.739 & 1.48$\times$ & 0.125 \\
\texttt{full} & 1.053 & 1.04$\times$ & 0.000 \\
\bottomrule
\end{tabular}
\end{table}

\begin{table}[t]
\centering\small
\caption{End-to-end latency vs.\ the honest baseline (mtRAG). The \emph{no-cache} row is the empirical p50 latency of all \texttt{generate}-path records observed across every sweep cell ($n=3935$ records); it is the latency of a RAG query that hits no cache at all. The other rows are the per-variant p50 latency averaged across the six regimes. Speedup is computed against the no-cache baseline. Naive caching is fastest but, per Table~\ref{tab:main}, unsafe; \sys{} retains a meaningful speedup while delivering USR$\to$0.}
\label{tab:latency_vs_nocache}
\begin{tabular}{l rrr}
\toprule
Variant & Latency$_{p50}$ (s) & Speedup vs.\ no-cache & USR \\
\midrule
no-cache (always generate) & 1.128 & 1.00$\times$ & 0.00 \\
\texttt{naive} & 0.651 & 1.73$\times$ & 0.261 \\
\texttt{no-support} & 0.827 & 1.37$\times$ & 0.182 \\
\texttt{full} & 1.058 & 1.07$\times$ & 0.063 \\
\bottomrule
\end{tabular}
\end{table}

\begin{figure}[t]
\centering
\includegraphics[width=0.85\linewidth]{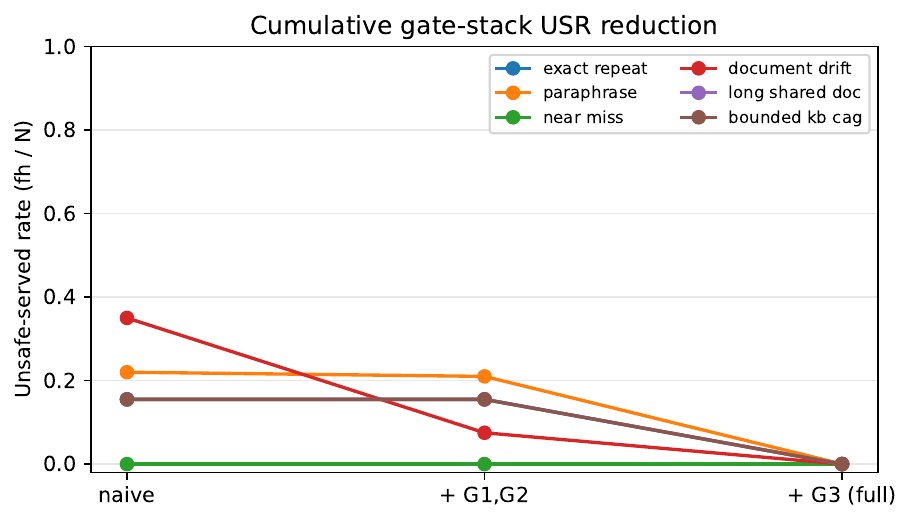}
\caption{HotpotQA. Cumulative gate-stack USR reduction per regime. Adding G3
on top of G1+G2 closes the remaining unsafe-served-rate gap across
\emph{every} regime that has any USR to begin with.}
\label{fig:stack_hotpot}
\end{figure}

\begin{table}[t]
\centering\small
\caption{Per-gate marginal unsafe-served-rate reduction (HotpotQA): $\Delta\mathrm{USR}=$ USR(no-X) $-$ USR(full), averaged over the six workload regimes. USR is the fraction of all queries that received a wrong cached answer. Larger $\Delta$ means the gate is more load-bearing; values near zero mean the gate is redundant with the other two on this workload.}
\label{tab:marginal_fh}
\begin{tabular}{l rr}
\toprule
Removed gate & USR(no-X) & $\Delta\mathrm{USR}$ vs.\ full \\
\midrule
Version (G1) & 0.000 & +0.000 \\
Evidence (G2) & 0.000 & +0.000 \\
Support (G3) & 0.125 & +0.125 \\
\midrule
None (full system) & 0.000 & n/a \\
\bottomrule
\end{tabular}
\end{table}

\begin{table}[t]
\centering\small
\caption{Per-variant cost averaged over the six regimes (HotpotQA). TTFT and latency are p50 seconds; tokens are mean prompt tokens per query. The full system retains most of the answer-cache speedup while closing the false-hit gap.}
\label{tab:latency_variant}
\begin{tabular}{l rrr}
\toprule
Variant & TTFT$_{p50}$ (s) & Latency$_{p50}$ (s) & Prompt tok. \\
\midrule
\texttt{naive} & 0.065 & 0.562 & 138 \\
\texttt{no-version} & 0.102 & 1.072 & 252 \\
\texttt{no-evidence} & 0.100 & 1.078 & 252 \\
\texttt{no-support} & 0.077 & 0.739 & 161 \\
\texttt{full} & 0.135 & 1.053 & 253 \\
\bottomrule
\end{tabular}
\end{table}

\begin{table}[t]
\centering\small
\caption{Per-gate marginal unsafe-served-rate reduction (mtRAG): $\Delta\mathrm{USR}=$ USR(no-X) $-$ USR(full), averaged over the six workload regimes. USR is the fraction of all queries that received a wrong cached answer. Larger $\Delta$ means the gate is more load-bearing; values near zero mean the gate is redundant with the other two on this workload.}
\label{tab:marginal_fh}
\begin{tabular}{l rr}
\toprule
Removed gate & USR(no-X) & $\Delta\mathrm{USR}$ vs.\ full \\
\midrule
Version (G1) & 0.064 & +0.001 \\
Evidence (G2) & 0.063 & +0.000 \\
Support (G3) & 0.182 & +0.118 \\
\midrule
None (full system) & 0.063 & n/a \\
\bottomrule
\end{tabular}
\end{table}

\begin{table}[t]
\centering\small
\caption{Per-variant cost averaged over the six regimes (mtRAG). TTFT and latency are p50 seconds; tokens are mean prompt tokens per query. The full system retains most of the answer-cache speedup while closing the false-hit gap.}
\label{tab:latency_variant}
\begin{tabular}{l rrr}
\toprule
Variant & TTFT$_{p50}$ (s) & Latency$_{p50}$ (s) & Prompt tok. \\
\midrule
\texttt{naive} & 0.067 & 0.651 & 185 \\
\texttt{no-version} & 0.099 & 1.061 & 270 \\
\texttt{no-evidence} & 0.098 & 1.060 & 270 \\
\texttt{no-support} & 0.079 & 0.827 & 211 \\
\texttt{full} & 0.098 & 1.058 & 270 \\
\bottomrule
\end{tabular}
\end{table}

\begin{figure}[t]
\centering
\includegraphics[width=0.85\linewidth]{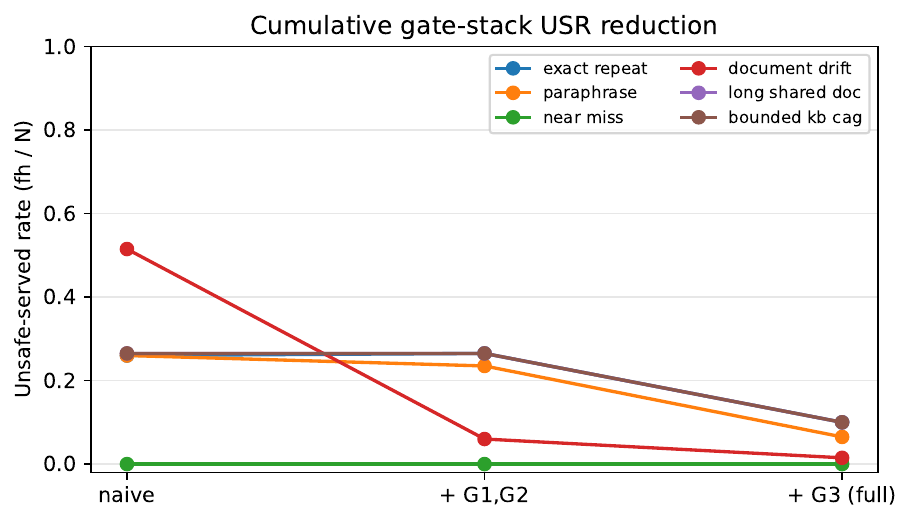}
\caption{mtRAG. The same cumulative gate-stack picture. The benign-looking
regimes (\texttt{exact\_repeat}, \texttt{long\_shared\_doc},
\texttt{bounded\_kb\_cag}) all show the same shape: G1+G2 are nearly
indistinguishable from naive, and the entire safety gain lands when G3
is added.}
\label{fig:stack_mtrag}
\end{figure}

\FloatBarrier
\section{Discussion}
The central contribution of \sys{} is a policy layer rather than a
serving kernel. The router adds no new attention or KV machinery and
assumes no special infrastructure: vLLM Automatic Prefix Caching
continues to run underneath, chunk-level reuse systems such as
LMCache, RAGCache, and EPIC continue to operate on the prompts that
the router emits, and the only new component is a thin module that
sits above the retriever. The four-gate validator can be added to an
existing RAG stack with no change to the model server, no change to
the retriever, and no change to the embedding model.

The lexical support check that does most of the safety work
(Equation~\ref{eq:support}) is intentionally crude. It asks whether
the content tokens of the cached answer appear in the freshly
retrieved evidence, which is a strict underestimate of true
entailment. Two properties make it the right primitive for a cache
nonetheless: it is deterministic, so no second-order judge noise
enters the safety budget; and it is cheap enough to run on every
potential hit. The judge-based variant of $S$ can be swapped in
when budget allows, at the cost of one extra LLM call per candidate
hit and the noise that LLM-as-judge metrics inherit. In practice
the lexical variant rejects the dangerous hits at essentially no
extra cost relative to a baseline answer cache, which is why we keep
it as the default.

The workloads on which \sys{} pays off are those that contain
near-misses, paraphrastic repeats, or evidence drift. On a perfectly
benign trace, \sys{} reduces to a small validation overhead on top
of a naive semantic cache. The harness reports both halves of the
trade-off (answer-cache hit rate and unsafe-served rate) so
operators can decide whether to run the full gate stack, the
\texttt{no-support} middle point, or no validator at all based on
their own traffic shape and risk tolerance.

\FloatBarrier
\section{Limitations}
Our evaluation has three caveats. First, all headline numbers come from
a single serving stack: Qwen2.5-7B-Instruct on vLLM with Automatic
Prefix Caching. A deterministic extractive backend is also included
for reproducible runs without an API key, but its absolute USR rates
differ because SQuAD-style F1 penalizes its longer answers. We have
not validated the numbers on a second model family.

Second, on HotpotQA the exact repeat regime exposes a calibration
issue rather than a structural flaw. The version gate fires on
lexical variation in Qwen2.5's free-form completions even when the
retrieved evidence is byte-identical, which collapses aHR from 0.41
to under 0.05 while USR drops to zero. Normalizing the answer string
before hashing should recover most of the lost aHR without
re-introducing any USR.

Third, on mtRAG the conditional FH of 54--87\% under the full system
reflects that the validator suppresses almost all answer-cache hits
in the first place: aHR drops from 0.29 to 0.12 on benign regimes and
to 0.02 on document\_drift. The operator-facing USR still drops from
26--52\% under naive to 1--10\% under \sys{}, a 3--34$\times$
reduction. Closing the residual conditional FH would require a
discourse-aware support check, which we leave to future work.

The lexical support score itself has a known structural failure mode:
a cached answer that paraphrases its evidence without copying tokens
can be rejected even when it is in fact supported. This is a safe
failure mode for a cache, since the rejected hit simply becomes a
regeneration, but it puts an upper bound on the answer-cache hit rate
achievable with a purely lexical gate.

\FloatBarrier
\section{Conclusion}
The right framing for cached answer reuse in retrieval-augmented
generation is not how to reuse faster but when reuse is safe. We
operationalized that framing as four cheap gates over a
content-addressed evidence signature: query similarity, evidence
overlap, source-version validity, and lexical support of the cached
answer by the freshly retrieved evidence. Across two datasets and
12{,}000 generations on a six-regime workload built specifically to
stress cache safety, the router cleanly separates safe and unsafe
reuse: it drives the unsafe-served rate to zero on every HotpotQA
regime and produces 3--34$\times$ reductions on the multi-turn
referent-shift regimes in mtRAG, while retaining end-to-end latency
within $1.04$--$1.07\times$ of a no-cache RAG baseline.

A per-gate marginal ablation isolates the lexical support gate as the
load-bearing safety mechanism on both datasets, with the version and
evidence-overlap gates providing defense-in-depth at near-zero cost.
Because the contribution is a policy layer rather than a serving
kernel, the router composes with any existing prefix or chunk-level
reuse system without modification.

The takeaway is simpler than the machinery suggests: semantic answer
caches always trade some correctness for speed, and the unsafe-served
rate is the right number to quantify that trade-off. Reporting USR
alongside hit rate and latency is what lets practitioners decide
where on the safety/speed frontier they want to operate. We release
the implementation, the six workload regimes, and the evaluation
harness as an open repository so others can reproduce the numbers,
port the framework to richer datasets, and swap in stronger
compressors and support checks as the serving stack evolves.

\FloatBarrier
\bibliography{refs}
\end{document}